\documentclass[12pt,preprint]{aastex}

\usepackage{graphicx}
\usepackage{epstopdf}
\usepackage{lscape}

\slugcomment{To appear in ApJ Letters}

\shorttitle{Torus in LLAGN}
\shortauthors{M\"uller S\'anchez et al.}

\begin{document}


\title{The central molecular gas structure in LINERs with low luminosity AGN: evidence for gradual disappearance of the torus\footnotemark[1]}


\author{F. M\"uller-S\'anchez$^{2,3,4}$, M.~A. Prieto$^{2,3}$, M. Mezcua$^{2,3,6}$, R. I. Davies$^{5}$, M. A. Malkan$^{4}$, M. Elitzur$^{7}$}

\affil{$^2$ Instituto de Astrof\'isica de Canarias, V\'ia L\'actea s/n, La Laguna, E-38205, Spain}

\affil{$^3$ Departamento de Astrof\'isica, Facultad de F\'isica, Universidad de la Laguna, Astrof\'isico Fco. S\'anchez s/n, La Laguna, E-38207, Spain}

\affil{$^4$ Department of Physics and Astronomy, University of California, Los Angeles, CA 90095-1547, USA}

\affil{$^5$ Max Planck Institut f\"ur Extraterrestrische Physik, Postfach 1312, 85741, Garching, Germany}

\affil{$^6$ Max Planck Institut f\"ur Radioastronomie, Auf dem Huegel 69, 53121, Bonn, Germany}

\affil{$^7$ Department of Physics \& Astronomy, University of Kentucky Lexington, KY 40506-0055, USA}

\footnotetext[1]{Based on observations at the European Southern Observatory VLT (082.B-0709 and 084.B-0568).}






\begin{abstract} 

We present observations of the molecular gas in the nuclear environment of three prototypical low luminosity AGN (LLAGN), based on
VLT/SINFONI AO-assisted integral-field spectroscopy of H$_2$ 1-0 S(1) emission at angular resolutions of $\sim0.17\arcsec$.
On scales of 50-150 pc the spatial distribution and kinematics of the molecular gas are consistent with a rotating thin disk, 
where the ratio of rotation ($V$) to dispersion ($\sigma$) exceeds unity. 
However, in the central 50 pc, the observations reveal a geometrically and optically thick structure of molecular gas ($V/\sigma<1$ and $N_H>10^{23}$ cm$^{-2}$) that is likely to be associated with the outer extent of any smaller scale obscuring structure. 
In contrast to Seyfert galaxies, the molecular gas in LLAGN has a $V/\sigma<1$ over an area that is $\sim9$ times smaller and column densities that are in average $\sim3$ times smaller. We interpret these results as evidence for a gradual disappearance of the nuclear obscuring structure. While a disk wind may not be able to maintain a thick rotating structure at these luminosities, inflow of material into the nuclear region could provide sufficient energy to sustain it. In this context, LLAGN may represent the final phase of accretion in current theories of torus evolution. While the inflow rate is considerable during the Seyfert phase, it is slowly decreasing, and the collisional disk is gradually transitioning to become geometrically thin. Furthermore, the nuclear region of these LLAGN is dominated by intermediate-age/old stellar populations (with little or no on-going star formation), consistent with a late stage of evolution.

\end{abstract}



\section{Introduction} \label{introduction}

For three decades, the ``Unified Model'' of Active Galactic Nuclei (AGN) has dominated discussion. Its most popular (and extreme) version assumes that AGN Types are distinguished 
only by viewing angle, with the broad lines in Type-2s blocked from our line-of-sight (LOS) by a highly-inclined molecular/dusty torus. 
This premise is supposed to hold for all luminosities, from low luminosity AGN (LLAGN) to powerful quasars (Antonucci 1993). 
More realistic formulations of the torus model involve additional variables such as covering factor or AGN luminosity (see the discussion on Elitzur 2012). 
Direct evidence for this model comes from spatially-resolved interferometric observations of thermal dust in a few prototypical AGN (e.g., Jaffe et al. 2004; Tristram et al. 2007). 
However, these observations do not provide sufficient information on the actual geometry (e.g., position-angle coverage necessary to constrain shape/orientation) and physical properties (e.g., dynamics) of the obscuring medium.   

By means of Adaptive Optics (AO) integral-field spectroscopy, Hicks et al. (2009, hereafter H09) showed that the molecular gas in the central tens of parsecs of Seyfert galaxies relates directly to the largest structures associated with the obscuring torus, 
as predicted by clumpy torus models (e.g., Nenkova et al. 2002, 2008; Schartmann et al. 2008): 
it is in a rotating disk-like distribution, 
has a high velocity dispersion relative to rotation ($V/\sigma<1$), and is optically thick (see also M\"uller-S\'anchez et al. 2006, 2009). 
HCN measurements of Seyfert galaxies suggest that the nuclear dense gas also has a large dispersion (Sani et al. 2012).
However, it remains unclear how this vertical structure can be supported. 
Some models suggest that the torus may be an evolving structure driven by accretion processes operating on scales
of tens of parsecs such as nuclear starbursts or disk-instabilities 
(e.g., Vollmer et al. 2008; Wada et al. 2009; Hopkins et al. 2012). When 
the inflow rate is low, 
the thickness and opacity of the torus decrease. 
Alternatively, by considering both the broad-line region and the torus to be 
different aspects of a disk-wind, \citet{elitzur09} argue that the torus thickness depends on AGN accretion. 
They predict that below a given luminosity threshold, 
a wind with a significant column density ($N_H$) can no longer be sustained. 
In either case, the geometrically and optically thick structure of molecular gas in LLAGN is expected to disappear. 

Observationally, there is an on-going debate on the existence of an obscuring structure in LLAGN\footnote{The bulk of the LLAGN population reside in low-ionization nuclear emission-line regions
(LINERs;  Ho 2008). Since the bolometric luminosity threshold
for separating low- and high-luminosity AGN is roughly defined as a few $10^{42}$ erg s$^{-1}$ (Ho 2009), 
some LLAGN reside in Seyfert galaxies. 
Strictly speaking, we study LINERs/LLAGN here, but refer to them just as LLAGN.}. 
The non-detection of an IR-bump in the SEDs of LLAGN \citep{jafo12} provides evidence for unobscured nuclei, as do the compact variable central UV sources observed in both types of LLAGN (e.g., Maoz et al. 2005). 
Furthermore, X-ray studies of nearby galaxies have found a correlation between $N_H$ and X-ray luminosity (e.g., Zhang et al. 2009). 
However, some LLAGN do exhibit high $N_H$ (e.g., NGC 3169; Terashima \& Wilson 2003). 
The detection of polarized broad-lines and ionization bicones in some LLAGN provides additional support for the presence of an obscuring torus (Barth et al. 1999; Pogge et al. 2000). 

In an effort to characterize for the first time the properties of the molecular gas in the central regions of LLAGN, 
we have observed three prototypical LINERs/LLAGN (Table 1) 
using AO-assisted $K$-band integral-field spectroscopy. 
The 2.12 $\mu$m  H$_2$ 1-0S(1) emission line is used to analyze the two-dimensional (2D) distribution and kinematics of the gas within the central 150 pc. 
These techniques have already been applied to study the properties of H$_2$ in NGC 1097 (H09; Davies et al. 2009). Here we bring those data together with new data on three additional objects (\S2 and \S3) to probe the putative obscuring medium in LLAGN.

\section{Observations and Data Processing}\label{observations}

The galaxies analyzed here were observed with AO and SINFONI at the VLT in two separate runs between March 2009 and March 2010. 
SINFONI provides spectra for every pixel in a contiguous 2D field
with $64\times32$ pixels \citep{eisenhauer03}. The pixel scale in all observations was $0.05\arcsec\times0.1\arcsec$, resulting in a field-of-view of $3.2\arcsec\times3.2\arcsec$. 
The spectral range was similar for all runs (1.95-2.45 $\mu$m) with typical spectral resolution R$\sim$4000 (FWHM$\sim70$ km s$^{-1}$).  For each galaxy, the nucleus was used as NGS by the AO-module. 
The exposure times for NGC 1052, NGC 2911, and NGC 3169 were 180, 20, and 60 minutes, respectively. The FWHM angular resolution, measured from the spatially unresolved non-stellar continuum in $K-$band, is $\sim0.17\arcsec$ in NGC 1052 and NGC 3169, corresponding to 15 and 16 pc, respectively, and $\sim0.23\arcsec$ (45 pc) in NGC 2911. 
Data reduction was performed using the SINFONI data reduction package SPRED \citep{abuter05}. 
Nearby standard stars (A- or B-type) were observed close in time to the science frames and used for telluric correction and flux calibration.
In addition, flux calibration was cross-checked with VLT/NACO and 2MASS photometric data in $1\arcsec-4\arcsec$ apertures. 

The resulting spectra show several emission lines of H$_2$ and stellar absorption features (Fig. 1). 
The 2D properties of H$_2$ 1-0 S(1) were extracted using the code LINEFIT (Davies et al. 2009). 
This procedure corrects automatically for instrumental broadening. Typical uncertainties in the kinematic measurements are between $5-20$ km s$^{-1}$. 

\section{Results and Analysis}\label{results} 

The SINFONI data reveal the morphology and kinematics of the molecular gas 
in unprecedented detail (Fig. 2; see H09 for the maps of NGC 1097). 
Similar morphologies are observed in the three sample galaxies. The flux distributions show bright central structures with elongations from Northeast to Southwest and morphologies suggesting filamentary formations. While in NGC 3169 (morphological type S), the photometric position-angle (PA) of H$_2$ is similar to the PA of the stellar continuum, in NGC~1052 and NGC~2911 (type E4 and S0, respectively), the PA of H$_2$ is strongly misaligned (by $\sim70-80\degr$) with that of the stars and the galactic major axis, indicative of an external origin of the gas, probably being acquired through merging, as has been proposed for elliptical galaxies \citep{morganti06}. 

The velocity fields of H$_2$ exhibit azimuthal symmetry, a zero-velocity axis (or kinematic minor axis) almost perpendicular to the kinematic major axis (which is also the photometric major axis in all cases), and small twists of the iso-velocity contours. 
Taken together, these features provide strong qualitative evidence for rotation in a disk, but with the presence of radial flows in the nuclear region.  
In NGC 1052 and NGC 2911, the gas has a kinematic major axis that is not aligned with the stellar rotation axis (Fig. 2), providing strong support to the hypothesis of an external origin of the molecular gas. 

While the H$_2$ velocity fields indicate rotating disks, the dispersion maps suggest that these disks are fairly thin in the outer $r>0.4\arcsec$ ($\sigma\sim20-30$ km s$^{-1}$), but remarkably thick in the vicinity of the AGN ($\sigma$ increases up to $\sim150$ km s$^{-1}$). In all cases, $\sigma$ also increases in a direction almost perpendicular to the H$_2$ kinematic major axis (in NGC 3169 this is observed only in the Southeast, Fig. 2). Our experience with previous integral-field observations of other galaxies, reveals that increases in dispersion in particular directions surrounding the AGN, are usually related to the presence of a passing radio-jet and/or outflows \citep{mueller11}. Radio and X-ray observations of NGC 1052 and NGC 3169 reveal extended emission in the direction of the enhancements in $\sigma$ \citep{kadler04, hummel87, terashima03}. A similar situation is observed in radio maps of NGC 2911 \citep{wrobel84, condon91}. Thus, the regions of enhanced $\sigma$-values (delineated by dashed lines in the dispersion maps, Fig. 2), are probably due to radio-jets or AGN-driven outflows. 

For quantitative modeling we adopted an 
axisymmetric rotating disk generated by the code DYSMAL (as in NGC 1097, Davies et al. 2009; 2011). 
For determining the disk parameters and their uncertainties, we adopted an exponential mass distribution of outer radius $1.6\arcsec$. We then varied inclination, PA, total mass, black hole (BH) mass and radial scale length ($R_{1/\mathrm{e}}$) to minimize $\chi^2$ in the 
residual velocity and dispersion maps. 
The regions perpendicular to the kinematic major axis containing high $\sigma$-values (the area inside the dashed lines in the dispersion maps, Fig. 2), which are likely of non-gravitational nature, were not included in the fit. 
The smearing due to the instrumental resolution is taken into account by convolving the inclined model with the Gaussian point-spread function of the appropriate width. 
Input errors in the $\chi^2$ evaluation were the $1\sigma$-fit uncertainties of the line profiles. All uncertainties given for the derived parameters are $68\%$ confidence. 

The best-fitting models are presented in Table 1 and illustrated with one-dimensional cuts along the kinematic major axis (Fig. 3). 
The models provide a good fit to the observed LOS-velocity and $\sigma$ curves, but the relatively large nuclear dispersion (at $r<0.4\arcsec$) cannot be accounted for by the models. The dominant kinematic component in all cases is well described by circular motion. However, the velocity residuals maps of NGC 1052 and NGC 2911 show non-circular velocities of up to $\sim60$ km s$^{-1}$ within $1.5\arcsec$ of the nucleus. 
Since the purpose of this investigation is to derive the best-fitting velocity field and study the bulk properties of H$_2$, we do not discuss the residuals here (including the regions where the outflows are observed). 
A detailed analysis of the kinematics
in these two galaxies will be presented in another publication. 
Overall, the observed properties of the nuclear H$_2$ in these LLAGNs indicate a disk-like distribution with significant rotation and high velocity dispersion in the central $r<0.4\arcsec$.

\section{Discussion}\label{discussion}

On scales of $r>50$ pc ($\sim100$ pc in NGC 2911) the properties of the molecular gas are consistent with a thin disk model (Fig. 3). 
Inside this radius, the dispersion increases faster than the models, and within $r<25$ pc ($\sim50$ pc in NGC 2911 but we use 25 pc for comparison purposes) 
the ratio $V/\sigma$ is $\leq1$, 
implying that the gas on these scales must be supported by rotation and random motions within a spheroidal structure (e.g., a geometrically thick disk). 
In addition, the gas in this region is optically thick. From the estimated dynamical masses, and assuming a typical gas fraction ($f_g$) in the centers of galaxies between $1-10\%$\footnote{The gas mass can also be estimated from the luminosity of 1-0 S(1) using an appropriate conversion factor \citep{mueller06}. We used the same factor as H09 of 430 M$_\sun$/L$_\sun$ (which gives $f_g\sim10\%$ for Seyferts), and derived an average $f_g\sim3.2\%$ in the central 50 pc of our sample of LLAGN. This is the value used throughout the paper. 
CO measurements of LLAGN also suggest $f_g$ of a few percent (Raluy et al. 1995; Tan et al. 2008).} 
(H09; Hopkins et al. 2012), we obtain an average $N_H$ of $1.7\times10^{23}$ cm$^{-2}$ (Table 1). In NGC 1052 and NGC 3169 the estimated $N_H$ is consistent with that derived from X-ray spectroscopy \citep{terashima03, kadler04}. 
Taken together, these characteristics suggest a rotating, geometrically and optically thick disk, and are consistent with those expected for the molecular torus in unification models.
A similar conclusion was reached for the nuclear H$_2$ in NGC 1097 (H09). 
Therefore, we refer to this structure as ``large-scale torus'', keeping in mind that 
the obscuring material probed by mid-IR interferometry of Seyferts has a size of a few pc or less 
(e.g., Jaffe et al. 2004), and that this is likely responsible for other functions such as collimation of ionization
cones \citep{mueller11}. Nevertheless, the contribution of the large-scale torus to AGN obscuration may be considerable, as suggested by current clumpy torus models  (e.g., Schartmann et al. 2008).

The large-scale torus in LLAGN appears to be different from the one in Seyfert galaxies. This conclusion is apparent from the ratio of azimuthally-averaged $V/\sigma$ as function of radius (Fig. 4). For comparison purposes, we also plotted the average of this function for six Seyfert  
and 
two starburst galaxies 
observed with SINFONI on the same spatial scales ($r\sim150$ pc). 
On average, for LLAGN, $V/\sigma$  is low ($0.8\pm0.3$) at $r=25$ pc and high ($4.6\pm1.2$) at 100 pc. Thus random motions are significant within $r<25$ pc, despite the rotation of the gas. In contrast, starburst galaxies exhibit a high $V/\sigma$ at any radius 
consistent with a thin disk (an almost uniform dispersion map with values $<30$ km s$^{-1}$, i.e. no significant thickening at the center), 
and Seyfert galaxies show low $V/\sigma$ at radii up to 100 pc, suggesting that 
the geometrically thick disk in these objects is more extended than in LLAGN. 
The obscuration in LLAGN is probably also lower than in Seyferts. 
The average 1-0 S(1) luminosity of the LLAGN in our sample within $r=25$ pc is $\sim1.5\times10^4$ $L_\sun$, approximately 3.2 times smaller than the average found for Seyfert galaxies (H09). If the H$_2$ luminosity reflects gas mass (rather than differences in excitation), then the LLAGN in our sample have gas masses (and in consequence $N_H$) on average $\sim3.2$ times smaller than Seyferts (H09). This is consistent with CO measurements of local AGN (including NGC 1097 and NGC 1052; Raluy et al. 1998; H09), which indicate that there is more gas in the centers of Seyferts than in LLAGN. 

Fig.~4 clearly shows that the molecular gas in LLAGN is in an intermediate state between an extended large-scale torus (a thick disk) and no large-scale torus at all (a thin disk). We interpret this structure as a large-scale torus in the process of gradual disappearance. This conclusion is supported by the theory of torus evolution proposed by Vollmer et al. (2008). In this model, a torus passes through 3 different phases dictated by an external mass accretion rate. Initiated by a massive gas infall, a turbulent thin disk is formed in which starbursts occur. Such disks, characterized by the production of several star clusters and high star formation rates ($SFR>10$ M$_\sun$ yr$^{-1}$), are observed in local starburst galaxies (M\"uller-S\'anchez et al 2010, Fig. 4). 
Once supernovae remove the intercloud medium, 
a collisional thick disk of dense gas clouds is formed with a decreasing, but still high accretion rate. The collisional thick disk also forms stars, but with an efficiency of $\sim10\%$ when compared to the previous stage ($SFR\sim1$ M$_\sun$yr$^{-1}$). The Seyferts in the SINFONI subsample (Fig. 4) are interpreted to be in this second phase (H09). The mass inflow rate is slowly decreasing, and the collisional disk is gradually transitioning to become geometrically thin. The $SFR$ reaches a minimum, and the nuclear region is dominated by intermediate-age/old stellar populations. Based on the observed properties, the LLAGN in our sample are in this transitional phase (Figs. 1 and 4). When the inflow rate has significantly decreased, the collisional torus either completely disappears (there is very little gas left) or becomes thin and transparent as the circumnuclear disk in the Galactic Center. 

The proposed scenario reveals a possible $Starburst\rightarrow Seyfert\rightarrow LINER/HII\rightarrow LINER\rightarrow Passive$ $Galaxies$ evolutionary sequence of AGN activity and BH growth. 
The initial phase of activity of a galactic nucleus is characterized by vigorous star formation. After $\sim10$ Myr, significant accretion onto the BH triggers the Seyfert phase. This highly active phase may be modulated by feeding and feedback processes. The final stages of accretion are observed in LINERs, in which massive BHs have already formed. 
This picture is supported by recent observational studies of local galactic nuclei (e.g., Constantin et al. 2009; Schawinski et al. 2010), and numerical simulations of the origin and evolution of luminous AGN (e.g., DiMatteo et al. 2005; Hopkins et al. 2006). 

The kinematic signatures of outflows detected in the dispersion maps suggest that a disk-wind may still be present. In the wind scenario, the main channel for release of energy may be switching at low luminosities from AGN-driven outflow to radio-jets. Thus, the regions in which we detect high $\sigma$ could be interpreted as gas from the ISM shocked by a radio-jet. However, it seems unlikely that this mechanism is also responsible for the high dispersion measured in the central 50 pc (at least not in its most part). First, the circular symmetry of these regions (extending out to $r\sim25-50$ pc) is not consistent with that expected from a collimated jet. In addition, in some objects, there is no clear evidence of a radio-jet (e.g., NGC 1097; Orienti \& Prieto 2010), and yet a low $V/\sigma$ 
is observed (Fig. 4). Although a contribution from the disk-wind to the large-scale torus seems unlikely, it could significantly contribute to sustain any smaller scale torus 
(the threshold for a $M_{\mathrm{BH}}=1\times10^8$ M$_\sun$ is $L_{\mathrm{bol}}\gtrsim2\times10^{40}$ erg s$^{-1}$; Elitzur \& Ho 2009). 
The dust sublimation radius in LLAGN with $L_{\mathrm{bol}}=10^{42}$ erg s$^{-1}$ is $r_{sub}=0.4(L_{\mathrm{bol}}/10^{45})^{1/2}$ pc $\sim0.012$ pc (Nenkova et al. 2008). The outer radius of the ``small-scale'' torus
is then $R=50-100r_{sub}=0.6-1.2$ pc (Nenkova et al. 2002; Tristram \& Schartmann 2011), which, at our resolution, is unresolved. 
These results suggest that the torus comprises at least two components on different scales each of which fulfills different roles. 
On scales of $\sim15$ pc and larger, our results reveal a diffuse and clumpy envelope 
which represents the reservoir of gas necessary to feed the AGN, and possibly contributes to nuclear obscuration. 
At smaller scales, IR interferometry provides evidence for a compact torus 
which is likely responsible for most of the obscuration and collimation. 

Our integral-field data confirm the presence of an obscuring structure of molecular gas in the central 50 parsecs of LLAGN, which indicates that in these objects there is potentially enough material to fuel the AGN.  
However, it appears to be disappearing as indicated by the lower column densities and sizes compared with those of Seyfert galaxies. Particularly interesting would be a study of the gas kinematics in AGN having lower luminosities and/or Eddington ratios than the ones presented here. In these objects either there is no detectable nuclear warm-H$_2$ structure (as in M87 and some other LLAGN observed with AO; Gebhardt et al. 2011, A. Barth private communication), or it is thin and transparent as predicted by the torus evolution model \citep{vollmer08}.



\acknowledgments

We thank the staff of ESO for their support, Aaron Barth for helpful discussions, and the anonymous referee for useful comments. 
We acknowledge financial support from CONACYT and {\it COST Action MP0905 Black Holes in a Violent Universe}.

\clearpage



\clearpage

\begin{figure}
\epsscale{.99}
\plotone{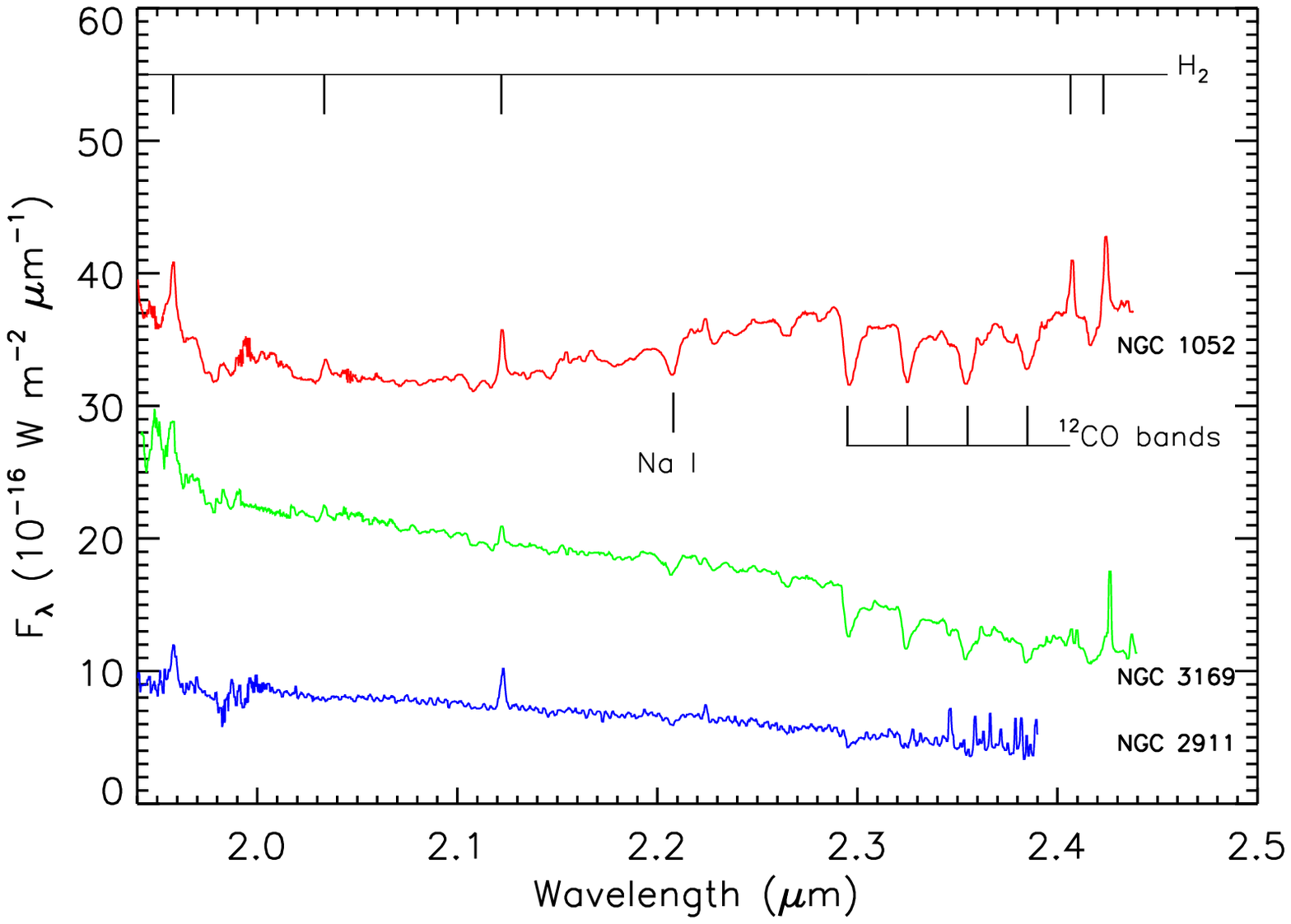}
\caption{$K-$band nuclear spectra of LLAGN extracted from our SINFONI data in $r=0.3\arcsec$ apertures. 
The prominent emission lines are from H$_2$ (marked at the top of the figure), the strongest being 1-0 S(1) at 2.12 $\mu$m. Several stellar absorption features (like the $^{12}$CO bandheads or NaI), characteristic of evolved stellar populations, are also noticeable. All $K-$band ionization lines (including Br$\gamma$)
are absent in these objects. 
NGC 1097 exhibits a similar spectrum, 
but Br$\gamma$ is marginally detected (Davies et al. 2007). 
Since Br$\gamma$ 
is normally considered a tracer of young stars, the marginally/non-detection of this line indicates little (as the very young and compact starburst in NGC 1097, Davies et al. 2007) or no ongoing star formation. 
\label{fig1}
}
\end{figure}

\clearpage

\begin{figure}
\epsscale{.99}
\plotone{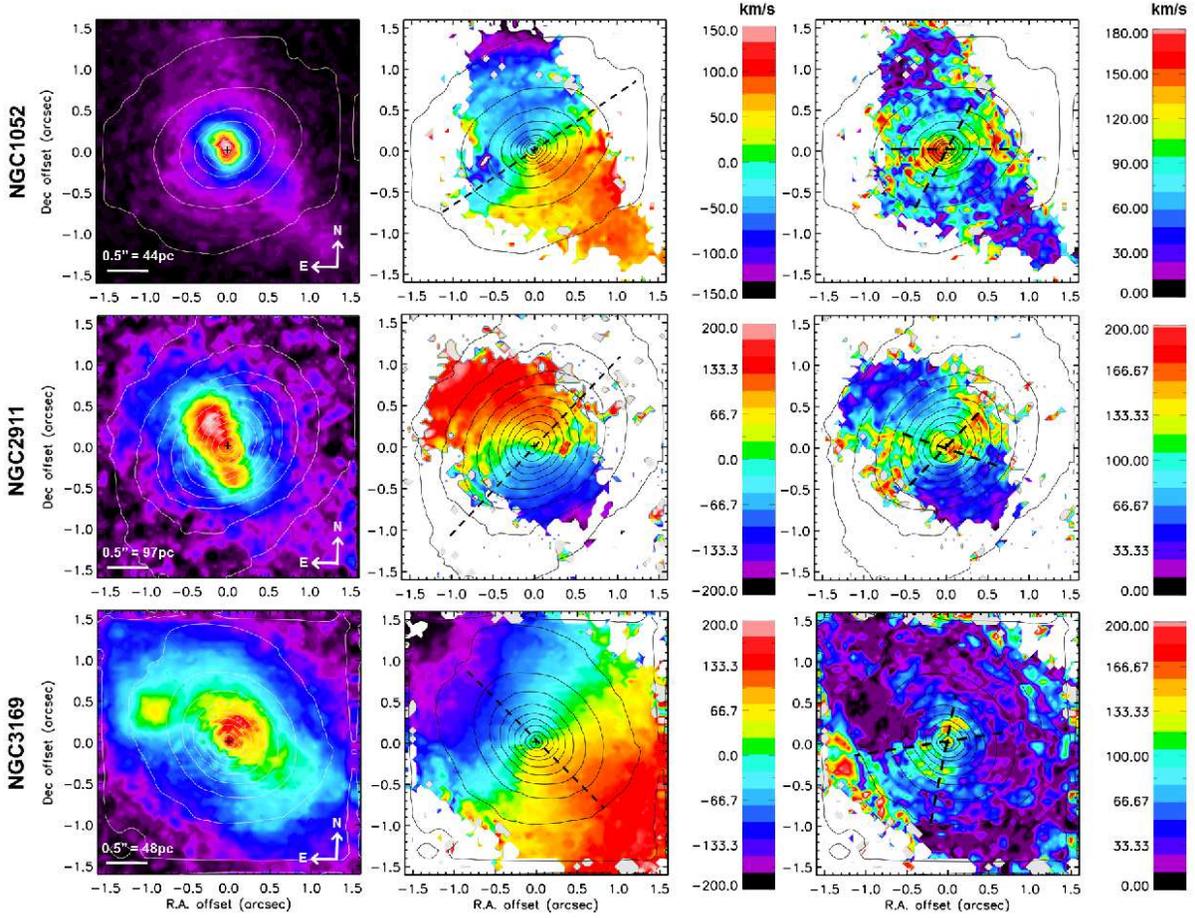}
\caption{Two-dimensional maps of H$_2$ 1-0 S(1) flux distribution, LOS-velocity, and velocity dispersion (from left to right), for each galaxy observed with SINFONI. 
Contours delineate the $K-$band continuum emission. The position of the AGN (the peak of non-stellar continuum at 2.2 $\mu$m) is marked with a cross. The dashed line in the middle column shows the orientation of the stellar kinematic major axis measured from the SINFONI data. In all cases this is consistent with the major axis of the galaxy. The two dashed
lines in the dispersion maps delineate approximately the regions where the radio and/or X-ray emission is observed in each galaxy. 
Rejected pixels in the velocity and dispersion maps are those with a flux density
lower than 5\% of the peak of H$_2$ emission.  
\label{fig2}}
\end{figure}

\clearpage

\begin{figure}
\epsscale{.5}
\plotone{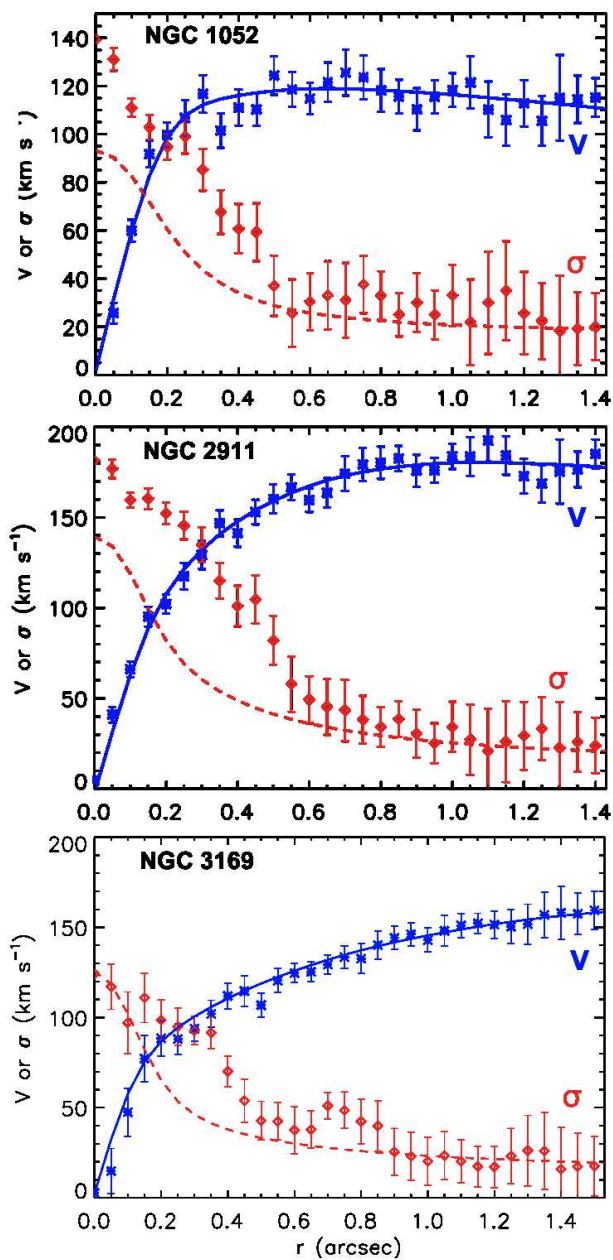}
\caption{Velocity and dispersion distributions of the molecular gas along the H$_2$ kinematic major axis of the sample galaxies (see H09 for the curves of NGC 1097). 
Asterisks represent LOS-velocity measurements and open diamonds dispersion values.  Continuous curves denote the best-fitting exponential disk model to the velocity datapoints, and dashed curves the same models but for $\sigma$. In all panels vertical error bars represent $1\sigma$ uncertainties in the measurements.  
\label{fig3}}
\end{figure}

\clearpage

\begin{figure}
\epsscale{.99}
\plotone{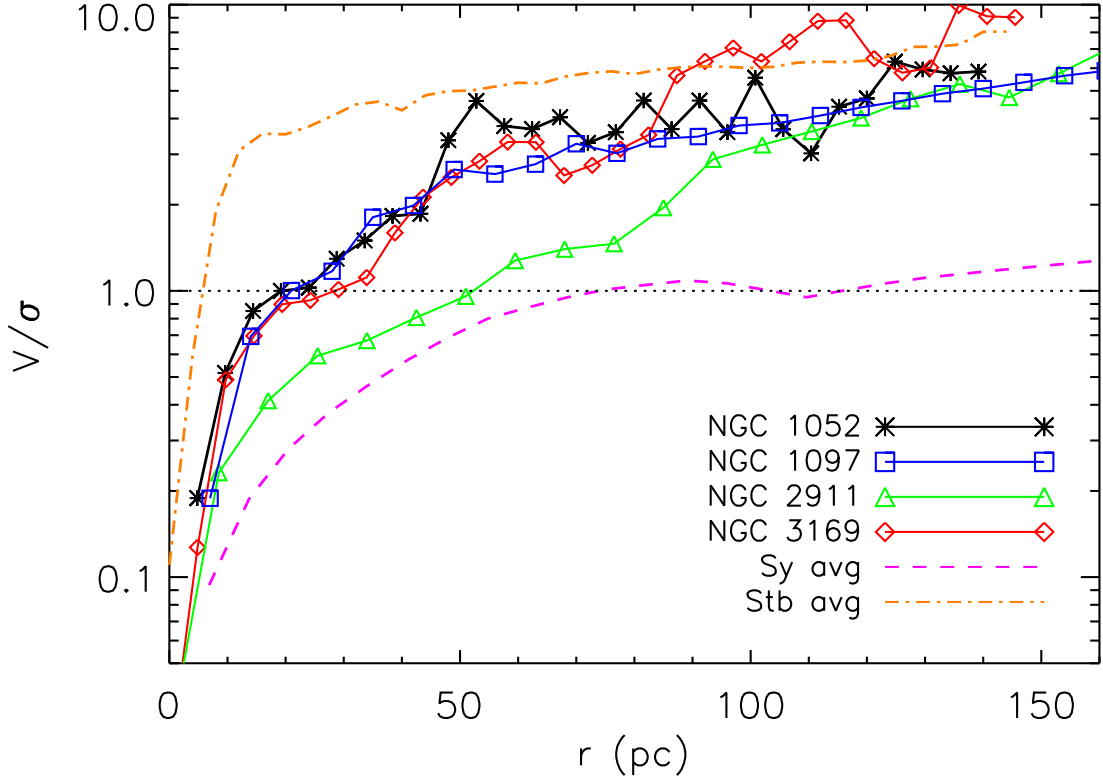}
\caption{Azimuthally-averaged $V/\sigma$ versus radius for each galaxy as indicated by the legend. 
The dotted horizontal line shows the point where $V/\sigma=1$. 
The typical error of the $V/\sigma$ estimates is $35\%$. The average $V/\sigma$ as a function of radius was obtained for Seyferts from the data of NGC 3783 and NGC 7469 (H09), NGC 2992 (Friedrich et al. 2010), NGC 6814 \citep{mueller11}, NGC 1566 and NGC 3081 (this work); and for starburst galaxies from NGC 253 and NGC 4945 (this work).
\label{fig4}}
\end{figure}

\clearpage

\begin{landscape}
\begin{table}
\begin{center}
{\scriptsize
\begin{tabular}{l c c c | c c c c c | c c c | c c c}
\hline
\hline \noalign{\smallskip}
\multicolumn{4}{c}{Basic data} \vline & \multicolumn{5}{c}{Global derived parameters} \vline & \multicolumn{3}{c}{At $r=25$ pc} \vline & \multicolumn{3}{c}{At $r=100$ pc} \\
Object & AGN & D\tablenotemark{a} & log$L_{\mathrm{bol}}$\tablenotemark{b} & 
PA & $i$ & $R_{1/\mathrm{e}}$ & log$M_{\mathrm{BH}}$\tablenotemark{c} & 
log$L_{\mathrm{bol}}/L_{\mathrm{Edd}}$ & 
$V/\sigma$ & log$M_{\mathrm{dyn}}$\tablenotemark{d} & log $N_{\mathrm{H}}$\tablenotemark{e} & 
$V/\sigma$ & log$M_{\mathrm{dyn}}$\tablenotemark{f} & log $N_{\mathrm{H}}$\tablenotemark{e} \\
  & type\tablenotemark{a} & (Mpc) & (erg s$^{-1}$) & ($\degr$) & ($\degr$) & (pc) & (M$_\odot$) & & & 
(M$_\odot$) & (cm$^{-2}$) &  & (M$_\odot$) & (cm$^{-2}$) \\
\hline \noalign{\smallskip}
NGC 1052 & L1.9 & 18 & 41.4 & $30\pm5$ & $55\pm4$ & $50\pm7$ & $8.0^{+0.2}_{-0.1}$ & $-4.9^{+0.2}_{-0.1}$ & 
$1.0\pm0.3$ & $8.3\pm0.1$ & $23.2\pm0.5$ & $4\pm1.0$ & 
$8.4^{+0.2}_{-0.1}$ & $22.1\pm0.5$ \\
NGC 2911 & L2 & 40 & 42.0 & $-156^{+7}_{-4}$ & $57\pm5$ & $170\pm15$ & 
$8.2^{+0.3}_{-0.2}$ & $-4.5^{+0.3}_{-0.2}$ & 
$0.6\pm0.2$ & $8.6\pm0.2$ & $23.5\pm0.5$ & $3\pm0.7$ & 
$9.0^{+0.2}_{-0.3}$ & $22.8\pm0.5$ \\
NGC 3169 & L2 & 20 & 42.4 & $47\pm3$ & $43\pm6$ & $120\pm10$ & $7.9^{+0.2}_{-0.1}$ & $-4.0^{+0.2}_{-0.1}$ & 
$0.9\pm0.3$ & $8.2\pm0.2$ & $23.1\pm0.5$ & $7\pm1.5$ & $8.7^{+0.3}_{-0.2}$ &  $22.5\pm0.5$  \\
 &  &  & & & & & & & & & & \\
NGC 1097 & L1 & 18 & 41.6 & $-52\pm5$ & $42\pm5$ & $ --- $ & $8.1^{+0.2}_{-0.2}$ & $-4.4^{+0.2}_{-0.2}$ & 
$1.2\pm0.4$ & $8.2\pm0.2$ & $23.0\pm0.5$ & $4\pm1.2$ & $8.9^{+0.2}_{-0.2}$ &  $22.6\pm0.5$  \\
\hline
\hline
\end{tabular}
}
\tablenotetext{a}{Spectroscopic classification and Distance (D) according to Ho et al. (1997), except for NGC 1097 (H09). L2: LINER without a broad H$\alpha$ line, L1.9: LINER with weak broad H$\alpha$ emission, L1: LINER with broad components of both H$\alpha$ and H$\beta$ lines. L2s in the sample show convincing evidence of being LLAGN. The two have a high brightness-temperature radio core, which is also a hard X-ray point source \citep{terashima03, wrobel84, ellis06}.} 
\tablenotetext{b}{Bolometric luminosity estimated using the X-ray luminosity as $L_{\mathrm{bol}} = 15L_{\mathrm{2-10 keV}}$ (Ho 2009), except for NGC 1097, which was estimated from the SED analysis (Prieto et al. 2010). X-ray luminosities taken from Ellis et al. (2006) and Terashima \& Wilson (2003).}
\tablenotetext{c}{Black Hole mass derived from the dynamical models, except for NGC 1097 (Davies et al. 2007).}
\tablenotetext{d}{Dynamical mass estimated taking into account the velocity dispersion of the gas as $M_{\mathrm{dyn}}=(V^2+3\sigma^2)r/G$, where $r$ is the radius from the nucleus and $G$ is the gravitational constant.}
\tablenotetext{e}{Hydrogen column density derived assuming the typical $1-10\%$ gas mass fraction (this interval corresponds to the lower and upper limits). Quoted values refer to $3.2\%$ gas fraction (see text for details).}
\tablenotetext{f}{Dynamical mass obtained from the kinematic modeling, except for NGC 1097 (H09). This is consistent with the estimate derived taking into account the velocity dispersion at this radius within the uncertainties.}
\end{center}
\caption[The Nearby AGN Sample]{Summary of basic data for LLAGNs and quantities derived from the H$_2$ 1-0 S(1) kinematics.
\tablecomments{Quoted uncertainties are 1 standard deviation of the fit, except for $N_{\mathrm{H}}$, which is delimited by the assumed gas fraction interval.}
\label{table1}}
\end{table}
\end{landscape}

\end{document}